\documentclass[epsf,prb,twocolumn,showpacs]{revtex4-1}

\usepackage[pdftex]{graphicx}
\usepackage{dcolumn}
\usepackage{bm}
\usepackage{epsfig}
\usepackage{latexsym}
\usepackage{amsmath}
\usepackage{amsfonts}
\usepackage{amssymb}
\usepackage{color}
\usepackage{array}
\usepackage{framed}

\begin{document}

\title{Nodal line entanglement entropy: \\
Generalized Widom formula from entanglement Hamiltonians}
\author{Michael Pretko \\ \emph{Department of Physics, Massachusetts Institute of Technology,
Cambridge, MA 02139, USA}}
\date{May 22, 2017}

\begin{abstract}
A system of fermions forming a Fermi surface exhibits a large degree of quantum entanglement, even in the absence of interactions.  In particular, the usual case of a codimension one Fermi surface leads to a logarithmic violation of the area law for entanglement entropy, as dictated by the Widom formula.  We here generalize this formula to the case of arbitrary codimension, which is of particular interest for nodal lines in three dimensions.  We first rederive the standard Widom formula by calculating an entanglement Hamiltonian for Fermi surface systems, obtained by repurposing a trick commonly applied to relativistic theories.  The entanglement Hamiltonian will take a local form in terms of a low-energy patch theory for the Fermi surface, though it is nonlocal with respect to the microscopic fermions.  This entanglement Hamiltonian can then be used to derive the entanglement entropy, yielding a result in agreement with the Widom formula.  The method is then generalized to arbitrary codimension.  For nodal lines, the area law is obeyed, and the magnitude of the coefficient for a particular partition is non-universal.  However, the coefficient has a universal dependence on the shape and orientation of the nodal line relative to the partitioning surface.  By comparing the relative magnitude of the area law for different partitioning cuts, entanglement entropy can be used as a tool for diagnosing the presence and shape of a nodal line in a ground state wavefunction.
\end{abstract}

\maketitle

\section{Introduction}

The concept of entanglement is a powerful tool for understanding, characterizing, and classifying quantum phases of matter.  For example, topologically ordered systems, such as gapped spin liquids, can be characterized by the inability of their spatial entanglement pattern to be smoothly deformed to a direct product state\cite{topo}.  Such highly-entangled phases of matter often have exotic excitations, such as anyons, and have been the focus of much research in recent years.  In order to characterize the entanglement in a many-body state, a useful prescription is to partition our system into two distinct spatial regions.  We can then form the reduced density matrix $\rho$ for a specific region by tracing out all degrees of freedom in its complement.  We then define the entanglement entropy of the partition as the von Neumann entropy associated with this density matrix, $S = -Tr[\rho\log\rho]$.  For ground states of local Hamiltonians in dimensions higher than one, the entanglement entropy typically obeys an area law\cite{area}, in that the entropy associated with a region of characteristic size $L$ in $d$ dimensions scales as the surface area, $S\propto L^{d-1}$.

A notable exception to this rule, however, occurs in Fermi surface systems.  For a garden variety codimension one Fermi surface, the entanglement entropy is known (both theoretically and numerically) to have a logarithmic violation of the area law, $S\propto L^{d-1}\log L$ .\cite{fermi1,fermi2}  The general intuition is that a Fermi surface is the result of sewing together many one-dimensional gapless theories, which also violate the area law logarithmically\cite{fermi3,anushya}.  To date, this is the largest violation of the area law in a well-established physical model.  (The same sort of logarithmic violation also occurs in systems with ``Bose surfaces"\cite{yang2}, where bosonic excitations have vanishing energy.  There are proposals for states with more severe violations\cite{ramis}, which may well be valid, but these results are less well understood.)  This is quite a striking result, since it can occur even in a free system (though the result also survives in an interacting Fermi liquid\cite{yang1}).  This is in contrast to most other highly entangled phases, which require strong interactions between the microscopic degrees of freedom.  The high degree of entanglement in these systems is facilitated by the large number of gapless modes, a very distinct feature of Fermi surfaces.  This logarithmic violation of the area law then serves as an important diagnostic tool for Fermi surfaces in a ground state wavefunction.  Obviously the result applies to free fermion systems, but it is arguably much more important in the study of strongly correlated systems, such as in composite Fermi liquids\cite{cfl} and spin liquids with spinon Fermi surfaces\cite{portal}, where the entanglement entropy can provide compelling evidence for emergent Fermi surfaces.  Furthermore, the coefficient of the $L^{d-1}\log L$ term in the entanglement entropy is universal, in that it depends only on the shape and size of the Fermi surface and the chosen partition, but does not depend on short-distance lattice physics.  This is in contrast to a standard area law, where the coefficient is highly sensitive to short-distance physics and can therefore be different in two systems with the same low-energy description.  The entanglement entropy (per spin species) in a Fermi surface system is given by the well-known Widom formula\cite{fermi2}:
\begin{equation}
S = \frac{\log L}{(2\pi)^{d-1}12}\int_{P.S.}\int_{F.S.}|\hat{n}_r\cdot\hat{n}_k|
\end{equation}
The two integrals are over the partitioning surface and the Fermi surface, respectively.  (Note that the integral over the partition provides the factor of area, $L^{d-1}$.)  The integrand is the inner product of the unit normals of the two surfaces, Fermi and partitioning.  This universal coefficient gives us important information about both the size and shape of the Fermi surface.  In particular, by choosing different partitions, we could determine any cross-sectional area of the Fermi surface and thereby mostly reconstruct the original shape.  (We note that Bose surface systems obey a similar, but slightly different Widom formula.\cite{yang3})

While the usual case of a codimension one Fermi surface is well-understood, comparatively little has been said about entanglement for Fermi surfaces of higher codimension, probably because there are far fewer examples that are not purely of academic interest.  The simplest case is a zero-dimensional (codimension $d$) Fermi surface, $i.e.$ a Fermi point, such as a Dirac cone.  In this case, for $d>1$, the entanglement entropy simply has an area law with a non-universal coefficient, and it is difficult to extract anything useful.  However, there is another important case which is a bit more interesting: a Fermi nodal line in three dimensions (codimension two).  As compared to a normal Fermi surface, a nodal line is harder to realize.  Naively, the simplest example would be a nodal line semimetal in a free fermion system.  Indeed, it is possible to have a stable band-touching along a line in momentum space, protected by symmetries\cite{balents,mullen,weng,liang,fang}.  However, it is not possible for this line to exist stably at the Fermi energy through any finite amount of symmetry protection\cite{balents,fstop}.  (Nevertheless, a Fermi surface which is only weakly deformed from such a line will behave roughly like a nodal line semimetal in certain parameter regimes, and recent experiments seem to indicate that the free fermion nodal line is of direct relevance for real materials\cite{hasan}.)  In actuality, the simplest realization of a nodal line occurs in certain superconductors, where there is no corresponding instability, and there is experimental evidence that such nodal lines occur in certain heavy-fermion superconductors\cite{hf1,hf2}.  Perhaps most interestingly, it has recently been shown that there exist stable nodal line spin liquids, consisting of a nodal line of fermionic spinons coupled to an emergent $U(1)$ gauge field\cite{yahui}.  Unlike their band-theoretic cousins, the nodal lines of these spin liquid states can be protected by symmetry to lie exactly at the Fermi energy.  In such a system, the entanglement entropy will have a large contribution from the many gapless modes along the nodal line (along with a separate gauge field contribution).  Finding the entanglement entropy for nodal lines is therefore relevant for a number of different physical systems.

In this work, we will find a generalized Widom formula which applies to Fermi surfaces of arbitrary codimension, with emphasis on the nodal line.  We will find that, unlike the codimension one case, higher codimension Fermi surfaces obey the area law.  For nodal lines in three dimensions (codimension two), we will find that the entanglement entropy is given by:
\begin{equation}
S = \frac{N}{a}\int_{P.S.} \int_{N.L.} |\hat{t}\times \hat{n}_r|^2
\label{nodal}
\end{equation}
where $\hat{t}$ is the unit tangent to the nodal line and $\hat{n}_r$ is the normal to the partitioning surface.  The integrals are over the partitioning surface and the nodal line, respectively.  $N$ is a non-universal numerical constant, and $a$ is the lattice cutoff, which we will address in a moment.  For general codimension $q$, the entanglement entropy becomes:
\begin{equation}
S_q = \frac{N_q}{a^{q-1}}\int_{P.S.}\int_{F.S.}|P_k \hat{n}_r|^q
\end{equation}
where $P_k\hat{n}_r$ is the projection of the partitioning surface normal $\hat{n}_r$ into the normal subspace of the Fermi surface, which we will explain in more detail later.  (This result also applies to the case of Fermi points, where the Fermi surface integral should be replaced by a discrete sum, and the projection operator becomes the identity.  The above equation then gives the standard area law.)

Note that the presence of the short-distance cutoff in these formulas makes the overall magnitude of the area law coefficient non-universal.  Nevertheless, we find an interesting universal dependence on the shape of the nodal line relative to the partitioning surface.  While the overall magnitude for a specific partition is not particularly meaningful, we can examine the ratio of the area law between different partitions, for which the non-universal prefactors will cancel out.  This will give us important information about the relative size of various cross-sections of the line.  The overall scale of the nodal line cannot be determined by this procedure and must be found through other means.  But once a scale is fixed, this formula allows one to map out the shape of the nodal line.  We will also discuss some of the unique features of Equation \ref{nodal} which can be used to distinguish nodal lines from other anisotropic entanglement patterns.

In order to find this generalized Widom formula, we must investigate the entanglement spectrum of Fermi surface systems.  The entanglement spectrum of a partition is defined as the spectrum of the reduced density matrix $\rho$ associated with a particular region.  This information can be nicely encoded in an object called the entanglement Hamiltonian $\tilde{H}$ (a.k.a. modular Hamiltonian), defined via:
\begin{equation}
\rho = e^{-\tilde{H}}
\end{equation}
Since $\rho$ is hermitian and has all non-negative eigenvalues, it is always guaranteed that we can write $\rho$ in this way.  In this sense, $\tilde{H} = -\log \rho$ is simply a trivial rewriting of the problem.  However, the usefulness of this form is the fact that, in many important cases, the entanglement Hamiltonian turns out to be local ($i.e.$ a sum of local operators).  For example, it is always local for any relativistic field theory partitioned into two half-spaces, as we will discuss in the next section.  By repurposing this relativistic tool, we will find that the entanglement Hamiltonian for Fermi surface systems, of any codimension, can also be written in local form in terms of a suitably defined set of low-energy variables, specifically a set of decoupled patch variables representing the different portions of the surface.  However, the result will still be nonlocal with respect to the original microscopic fermions, in agreement with earlier treatments of free fermion entanglement Hamiltonians\cite{peschel}.

\section{A Relativistic Trick}

In order to find the entanglement entropy for Fermi surface systems, we will repurpose a trick from the high energy playbook.  Consider the ground state of any relativistic field theory, and let us choose our partition to be into two half spaces, $x_1<0$ and $x_1>0$.  Let the Hamiltonian density of our theory be $\mathcal{H}$, such that the Hamiltonian is $H = \int d^dx \mathcal{H}$.  After tracing out the half-space $x_1<0$, the entanglement Hamiltonian of the ground state in the remaining region takes the form\cite{wichmann,bianchi}:
\begin{equation}
\tilde{H} = \int_{x_1>0} d^dx \, \bigg(\frac{2\pi x_1}{c}\bigg)\mathcal{H}
\label{bw}
\end{equation}
This result is sometimes known as the Bisognano-Wichmann (BW) theorem.  (Note that we will keep the speed of light $c$ explicit throughout, for reasons which will be apparent later.  Other constants, like $\hbar$ and $k_B$, will always be taken to be unity.)  A short version of the proof of this statement is given in Appendix A.  The intuition is fairly simple and is closely related to the fact that $\tilde{H}$ is the generator of boosts in the $x_1$ direction.  Notably, the proof only relies on a Lorentz symmetry between the $x_0$ (time) and $x_1$ coordinates.  The other spatial coordinates can simply be regarded as parameters and come along for the ride.

The result for a planar partitioning surface may seem like a special case, but it is not.  In fact, this example is sufficient to provide a general understanding of the area law for entanglement entropy \cite{bianchi,grover}.  For the ground state of a local Hamiltonian, the largest part of the entanglement is generally between degrees of freedom separated by very short distances, since they are directly coupled.  This is manifest from the fact that the area law coefficient typically diverges as the lattice scale $a$ is taken to $0$, indicating that most of the entanglement is coming from short-distance physics.  Thus, in order to get the leading contribution to the entanglement entropy, we only need to consider the region in the immediate vicinity of the partitioning surface.  In order to get important subleading corrections, such as topological entanglement entropy, one needs to be a bit more careful.  But for the dominant term of the entanglement entropy, which is what we will concern ourselves with here, we only need to worry about local properties of the partitioning surface.

\begin{figure}[t!]
 \centering
 \includegraphics[scale=0.3]{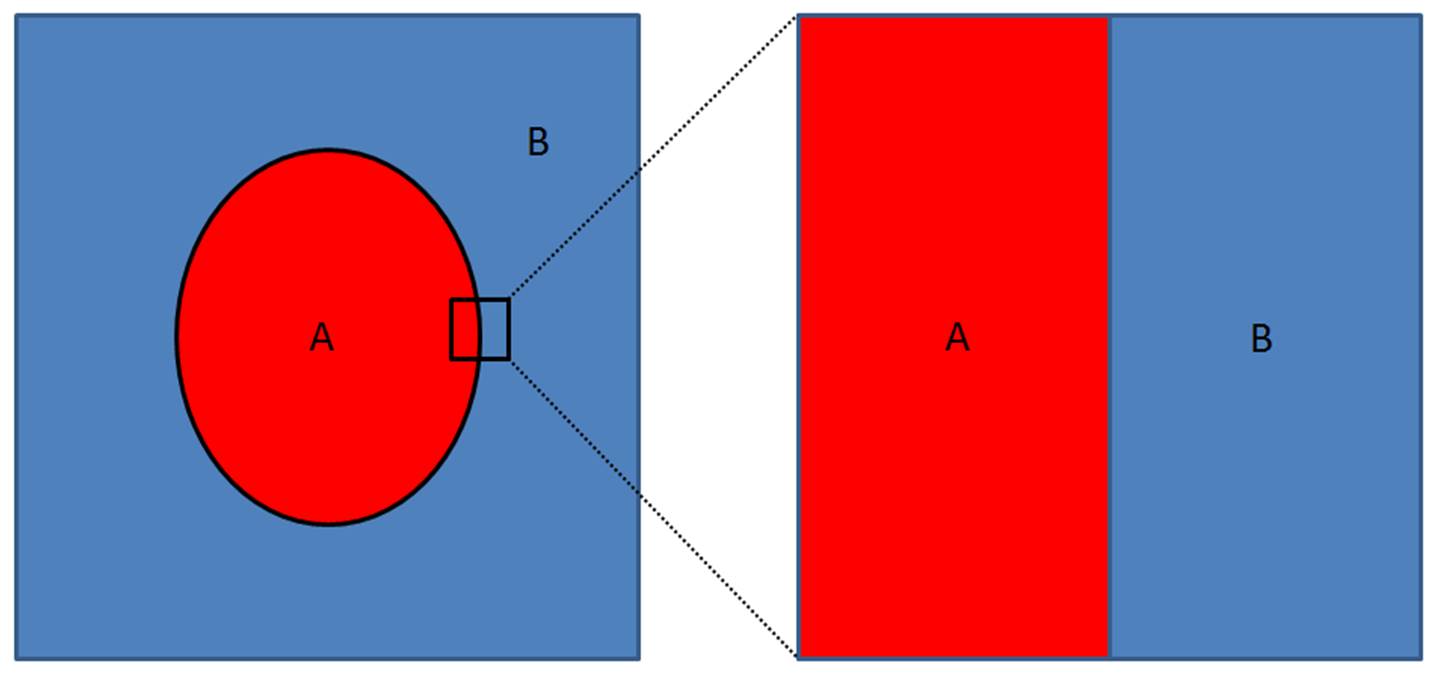}
 \caption{On scales small compared with $L$, the partitioning surface looks planar.}
 \label{fig:planar}
 \end{figure}

Using this intuition, instead of viewing the partitioning surface globally, we zoom in on a small patch.  We choose the dimensions of the patch to be large compared to the lattice scale $a$, so that the continuum description is valid, but small compared to the size of the partition $L$, so that the precise shape of the surface is irrelevant.  On this scale, the curvature is insignificant and the surface looks flat (see Figure \ref{fig:planar}), corresponding to the planar surface considered earlier.  Since the dominant contribution to entanglement is local ($i.e.$ intrapatch, not interpatch), we can evaluate the corresponding contribution to the entanglement entropy from one patch, then add up the contributions from each such patch to get the total entanglement entropy.  From this perspective, the area law seems inevitable, coming from integrating over all patches which make up the partitioning surface.  The entanglement entropy $must$ contain a factor of area, and it is the deviations from the area law which seem more difficult to explain.  In order to see how these violations arise, we have to develop a method for actually extracting the entanglement entropy from the entanglement Hamiltonian.

To do so, we must take a closer look at our reduced density matrix:
\begin{equation}
\rho\propto \exp\bigg\{-\int_{x_1>0}d^dx\bigg(\frac{2\pi x_1}{c}\bigg)\mathcal{H}\bigg\}
\end{equation}
We note that this density matrix looks very much like a thermal density matrix, $e^{-\beta H}$, except that now the temperature depends on position.  We write our density matrix as:
\begin{equation}
\rho \propto \exp\bigg\{-\int \beta(x)\mathcal{H}(x)\bigg\}
\end{equation}
where our locally defined temperature is $T(x) = \beta^{-1}(x) = \frac{c}{2\pi x_1}$.  Essentially, the problem of ground state entanglement has mapped onto a thermodynamic problem, one where the partitioning surface is incredibly hot, and the temperature falls off to zero in the bulk.  This makes some intuitive sense, in that far away from the surface the ground state is unaffected by the tracing out procedure, due to the primarily local nature of the entanglement.

With this interpretation in hand, we can then find the dominant contribution to the entanglement entropy by summing up the local thermal entropy densities:
\begin{equation}
S_{ent} = \int_{x_1>0}d^dx\,\, s_{therm}(T(x))
\label{thermal}
\end{equation}
This local thermal framework is reminiscent of a Thomas-Fermi sort of approximation and should be sufficient for extracting the leading piece of the entanglement entropy.  To demonstrate the power and the essential correctness of this thermodynamic method, let's show that it reproduces the exact universal result for a one-dimensional conformal field theory, as outlined in \cite{temperature}.  To avoid confusion with the speed of light $c$, we will denote central charges by $C$.  In a relativistic CFT, the thermal entropy density is given by \cite{fradkin}:
\begin{equation}
s_{therm}(T) = \frac{\pi C}{3c} T
\end{equation}
Let us take our partition to be a finite segment of length $L$.  The integral over area then becomes a sum over the two endpoints of the segment.  Using our prescription of integrating the thermal entropy over the half-space, we obtain the entanglement entropy as:
\begin{align}
S &= 2 \int_{x_1>0} dx_1 \frac{\pi C}{3c}\frac{c}{2\pi x_1} = \\
 \frac{C}{3} &\int_{x_1>0} dx_1\,\frac{1}{x_1} = \frac{C}{3} \log(L/a)
\label{1dthermal}
\end{align}
where we have taken a short-distance cutoff at the lattice scale, $a$, and a large-distance cutoff at the size of the interval, $L$.  Notice that the speed of light has completely canceled out.  This is the exact result known for the entanglement entropy of finite segments in conformal field theories \cite{cardy}.  Furthermore, it readily generalizes to the case where the region in question is a disjoint set of intervals, instead of just a single interval.  The fact that these equations match exactly, not just in terms of dependence on parameters but in numerical coefficient as well, is strong evidence that this method can be taken seriously.

Notice that the result for a one-dimensional CFT violates the area law (which in one dimension would simply be $S = $ constant) by a logarithmic factor.  This logarithmic divergence comes about from the large $x_1$ behavior of the integral in Equation \ref{1dthermal}, which corresponds to the low-temperature thermal behavior $s(T)\sim T$ in one dimension.  Such connections between entanglement and thermodynamics have been noted before\cite{swingle,mcgreevy}.  In higher-dimensional CFTs, we would have $s(T)\sim T^d$.  The resulting thermal integral will then converge at large $x_1$, resulting in an area law.  More generally, we can say for a relativistic system that whether or not the area law is obeyed is a direct consequence of the low-temperature thermal entropy behavior.  Suppose the low-temperature thermal entropy obeys a power-law, $s_{therm}(T)\sim T^\eta$.  For $\eta>1$, the thermal integral of Equation \ref{thermal} will converge at large $x_1$, giving an area law.  For $\eta = 1$, such as in one-dimensional CFTs, the area law is violated logarithmically.  (We will see that similar logic holds for Fermi surfaces.)  For $\eta<1$, there should be power law violations of the area law, $S\sim L^{d-\eta}$.  However, it seems unlikely that one can engineer a relativistic system to have such a large low-temperature thermal entropy.  If it is gapless, then it is likely described by a CFT, which always has $\eta \geq 1$.  If it is gapped, then the thermal entropy is exponentially small at low temperatures.  Thus, logarithmic violations of the area law are likely the most severe type of violation possible in a relativistic system.

\section{Application to Fermi Surface Systems}

For relativistic theories, we have seen that the entanglement Hamiltonian is relatively simple.  In a condensed matter context, however, we are typically concerned with explicitly nonrelativistic systems, where exact results are difficult to come by.  One important result is that, for a quadratic Hamiltonian, the entanglement Hamiltonian is also quadratic\cite{peschel}.  For example, in a system of fermions freely hopping on a lattice, $H = \sum_{ij}t_{ij}c^\dagger_i c_j$, the entanglement Hamiltonian must take the similar form:
\begin{equation}
\tilde{H} = \sum_{ij}h_{ij}c^\dagger_i c_j
\end{equation}
where $c^\dagger$ and $c$ are the fermion creation and annihilation operators and the sum runs over pairs of lattice sites.  Formally, this looks just like another free fermion Hamiltonian.  Unfortunately, the matrix elements $h_{ij}$ are in general neither translationally invariant nor even local.  Except in certain special cases, there is no closed form solution for the matrix elements, and they can typically only be determined numerically.  While in principle we know that such a quadratic result must hold, in practice it is difficult for us to directly extract the entropy from this form of the entanglement Hamiltonian.  (The quadratic result has been useful, however, in the study of entanglement in topological insulators and superconductors\cite{topins}.)  In Appendix B, we will show how to make connection with this microscopic result, but for now, we will take a different approach.
 
\begin{figure}[t!]
 \centering
 \includegraphics[scale=0.42]{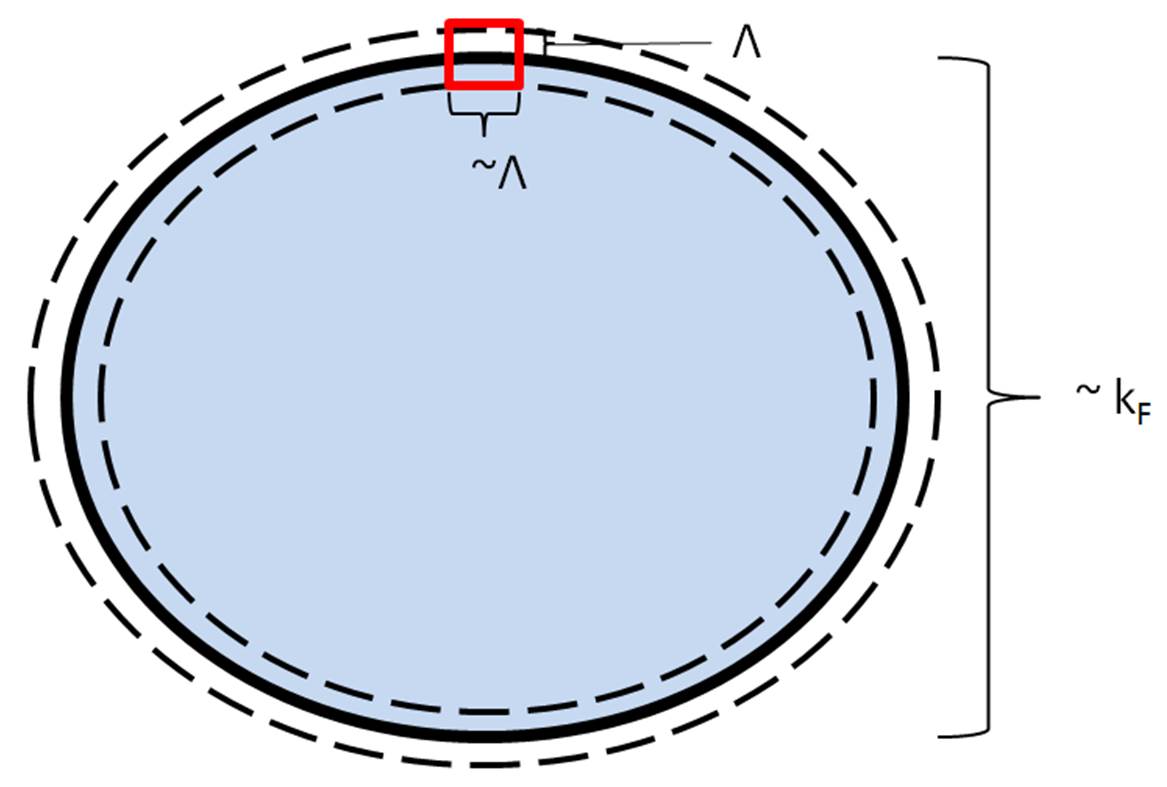}
 \caption{Though our Fermi surface has size of order $k_F$, all low-energy degrees of freedom lie within order $\Lambda$ of the Fermi surface.  For our arguments, we look at individual patches, such as the red box shown above, which is characterized by size $\Lambda$ in all directions.}
 \label{fig:patches}
 \end{figure}

In order to proceed, we shall now adapt the BW result from the previous section to obtain the entanglement Hamiltonian for a Fermi surface system with a spatial partition into two half-spaces.  We will then obtain the entanglement entropy for a generic partition by integrating over the partitioning surface, as before.  To accomplish these goals, we will work with an effective low-energy patch theory of the Fermi surface, which will essentially be equivalent to a set of lower-dimensional relativistic theories.  We will then apply the BW result (with subtle modifications) separately to each patch.  To realize this explicitly, we shall break up a Fermi surface of characteristic size $k_F$ into small patches of size $\Lambda \ll k_F$, as depicted in Figure \ref{fig:patches}.  All modes outside these patches have already been integrated out of our effective low-energy theory.  Each patch of the Fermi surface (labeled by index $p$) will be described by its own independent Hamiltonian, $H = \sum_p H_p$.  

The Bisognano-Wichmann result discussed above is only strictly applicable to theories with relativistic invariance.  However, the coordinates parallel to the partitioning surface are irrelevant to the proof.  We just need a spacetime symmetry between the one perpendicular spatial coordinate ($x_1$) and time.  Importantly, the effective temperature of the local thermal distribution is determined by the appropriate ``speed of light" of the system.  The correct velocity to use is technically the velocity characterizing the spacetime symmetry between $x_0$ and $x_1$.  For a strictly relativistic system, this point would be irrelevant since the velocity is the same in all directions, but this distinction will be crucial in our application to Fermi surfaces.

\subsection{Codimension One}

\begin{figure}[b!]
 \centering
 \includegraphics[scale=0.55]{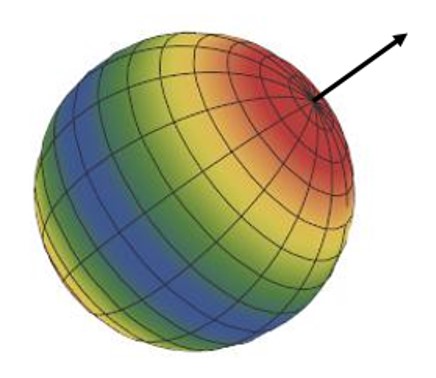}
 \caption{A map of the effective temperature on a spherical Fermi surface.  The surface is hottest (red) along the poles determined by the direction of $\hat{n}_r$, cooling off to zero temperature (blue) along the equator.}
 \label{fig:spheretemp}
 \end{figure}

As a warmup, we will find the entanglement Hamiltonian for a codimension one Fermi surface, then use it to rederive the Widom formula.  We start with a bare bones model of a codimension one Fermi surface of spin-$1/2$ fermions characterized by a Fermi velocity $v_F(k)$, which in general may vary over the Fermi surface.\cite{foot2}  We then break the Fermi surface up into patches, as in Figure \ref{fig:patches}, chosen small enough that $v_F$ is constant within each patch and we can view the surface as locally flat.  Each such patch on the surface essentially represents an independent chiral one-dimensional mode propagating in the normal direction with linear dispersion.  We can then write a Hamiltonian for the system as follows\cite{foot3}:
\begin{equation}
H = \sum_{\textrm{patches}}\int_{|k| < \Lambda} d^dk\, v_F\, \overline{\Psi}_p(k_\perp \sigma_x)\Psi_p
\end{equation}
We have an independent Weyl fermion field $\Psi_p$ for each patch, labeled by index $p$.  The variable $k_\perp$ represents the normal direction to the Fermi surface (which is different at each patch), and we assume that all modes with $k_\perp > \Lambda$ have already been integrated out of the problem.  The Fermi velocity depends on the patch, but is a constant within a specific patch.  Letting $x_\perp$ be the spatial coordinate corresponding to $k_\perp$, we can perform a Fourier transform within each patch (cutting off all $k$ integrals at $\Lambda$) and write the Hamiltonian in the following real-space form:
\begin{align}
\begin{split}
H = \sum_{\textrm{patches}} \int d^d x \,\,& v_F \overline{\Psi}_p (\sigma_x\partial_{x_\perp}) \Psi_p \equiv \\ \sum_{\textrm{patches}} &\int d^dx \mathcal{H}_p
\end{split}
\end{align}
Importantly, note that the real space field $\Psi_p(x)$ is not the same as the original microscopic fermion field, which would have involved a global Fourier transform, as opposed to a patchwise Fourier transform.  The corresponding Lagrangian is:
\begin{align}
\begin{split}
L = \sum_{\textrm{patches}} \int d^d x&\,\, \overline{\Psi}_p(\partial_t - v_F\sigma_x\partial_{x_\perp})\Psi_p = \\
\sum_{\textrm{patches}}&\int d^dx\,\,\overline{\Psi}_p(\sigma\cdot\partial)\Psi_p
\end{split}
\end{align}
where $\sigma \equiv \{1,\sigma_x,0,0,...\}$.  We now have a set of independent relativistic-looking theories for each patch of the Fermi surface.  However, we must be careful in applying our relativistic results.  As we have seen before, using the BW theorem amounts to a spacetime rotational symmetry between the time direction and the $x_1$ direction (partitioning surface normal), not the $x_\perp$ direction (Fermi surface normal).  In order to use the BW result, we must first rewrite the Lagrangian as:
\begin{align}
\begin{split}
L = &\sum_{\textrm{patches}} \int d^d x\,\,\bigg( \\
&\overline{\Psi}_p(\partial_t - v_F\sigma_x\cos\theta \partial_{x_1} -v_F\sigma_x\sin\theta\partial_{x_2})\Psi_p\bigg)
\end{split}
\end{align}
where $\cos\theta = |\hat{n}_p\cdot \hat{x}_1|$ represents the angle between the Fermi surface normal $\hat{n}_p$ (at the given patch) and the partitioning surface normal $\hat{x}_1$.  The coordinate $x_2$ is an appropriately chosen orthogonal coordinate.  We can now regard $\{x_2,...,x_d\}$ as parameters, viewing the above Lagrangian as a sum of one-dimensional theories with velocity $v_F \cos\theta$.  We can therefore simply quote the BW result for a given patch, provided we use the velocity $v_F \cos\theta$ as the ``speed of light."  Our entanglement Hamiltonian then becomes:
\begin{align}
\begin{split}
\tilde{H} = \sum_{\textrm{patches}} \int_{x_1>0} d^dx\, \frac{2\pi x_1}{v_F|\hat{n}_p\cdot \hat{x}_1|} \mathcal{H}_p = \\
\sum_{\textrm{patches}} \int_{x_1>0} d^dx\,\, \frac{2\pi x_1}{|\hat{n}_p\cdot \hat{x}_1|} \overline{\Psi}_p(\sigma_x \partial_{x_\perp})\Psi_p
\label{eh}
\end{split}
\end{align}
Note the presence of two different non-orthogonal coordinates in the last formula: $x_1$ represents the coordinate along the normal to the entangling surface, and $x_\perp$ represents the (spatial) coordinate along the direction of the normal to the Fermi surface.  As in our earlier analysis, we can interpret the resulting density matrix, $\rho = e^{-\tilde{H}}$, as a local thermal ensemble, with position dependent temperature.  The main difference now is that the temperature depends not only on spatial location but also on the patch considered, so each patch of the Fermi surface is at its own independent temperature given by:
\begin{equation}
T_p(x) = \frac{v_F|\hat{n}_p\cdot \hat{x}_1|}{2\pi x_1}
\label{temp}
\end{equation}
(Recall that $v_F$ in general can vary from one patch to another.)  Patches of the Fermi surface which align with the partitioning surface have the highest temperature and will contribute the most to the entanglement entropy.  At the other extreme, patches orthogonal to the partitioning surface have zero temperature and make no contribution to the entanglement entropy whatsoever.  This temperature profile is illustrated in Figure \ref{fig:spheretemp}.

In order to obtain the entanglement entropy, we will once again invoke our procedure of integrating the local thermal density, $S_{ent} = \int_{x_1>0} s_{therm}(T(x))$.  This amounts to a fairly simple thermodynamics problem.  We recall that the density of states (per real space volume and per spin species) contributed by a small patch of the Fermi surface of area $d\mathcal{A}_k$ is given by:
\begin{equation}
d\rho = \frac{d\mathcal{A}_k}{(2\pi)^d v_F(k)}
\end{equation}
and its contribution to the low-temperature entropy density is given by:
\begin{equation}
ds = \frac{\pi^2}{3}d\rho\, T = \frac{\pi^2 T}{3(2\pi)^d v_F(k)} d\mathcal{A}_k
\label{entdens}
\end{equation}
Using the temperature found in Equation \ref{temp}, we then find that the total entanglement entropy for the planar partitioning surface is given by:
\begin{align}
\begin{split}
S = \int_{F.S.}\int_{x_1>0} d^dx \frac{\pi^2}{3(2\pi)^d}\frac{v_F|\hat{n}_p\cdot \hat{x}_1|}{v_F(2\pi x_1)} = \\
\frac{\log L}{(2\pi)^{d-1} 12} \int_{F.S.}\int dx_2...dx_d \,(|\hat{n}_p\cdot \hat{x}_1|)
\end{split}
\end{align}
where the leading integral is over the Fermi surface.  The remaining spatial integrals represent an integral over the planar partitioning surface.  For a more general partitioning surface, we simply break it up into small patches, which locally look flat.  The same formula will apply, replacing $\hat{x}_1$ by the appropriate normal direction.  For a general partitioning surface, the entanglement entropy (per spin species) is then given by:
\begin{equation}
S = \frac{\log L}{(2\pi)^{d-1} 12} \int_{F.S.}\int_{P.S.} |n_r\cdot n_k|
\end{equation}
where the integrals are over the Fermi surface and the partitioning surface, which have normals $\hat{n}_k$ and $\hat{n}_r$ respectively.  This is precisely the Widom formula for the entanglement entropy of a Fermi surface \cite{fermi2}.  It is interesting to note that the Fermi velocity cancels out on each patch of the Fermi surface, so the entanglement entropy depends only on the geometry of the Fermi surface, not on the details of the dispersion.

Before moving on, we pause to put this equation into a useful alternative form.  We note that multiplying the Fermi surface area element by the factor $|\hat{n}_r\cdot \hat{n}_k|$ gives the area of the projection of that patch onto the partitioning surface.  Taking a convex Fermi surface for simplicity, it is not hard to see that integrating this projected area over the whole Fermi surface will give twice the extremal cross-section of the Fermi surface, as seen from the $\hat{n}_r$ direction (in other words, the size of its shadow).  We shall denote this cross-section by $\sigma_k(\hat{n}_r)$.  Then the Widom formula becomes:
\begin{equation}
S = \frac{\log L}{(2\pi)^{d-1}6} \int_{P.S.} \sigma_k(\hat{n}_r)
\label{cross}
\end{equation}
The Widom formula thereby allows us to obtain direct useful information about the size and shape of the Fermi surface.  In particular, choosing a planar partitioning surface in the $\hat{n}_r$ direction will directly tell us the cross-section of the Fermi surface in that direction.  The Widom formula therefore allows us to (almost) completely map out the shape of the Fermi surface using only the ground state entanglement entropy.

\begin{figure}[t!]
 \centering
 \includegraphics[scale=0.4]{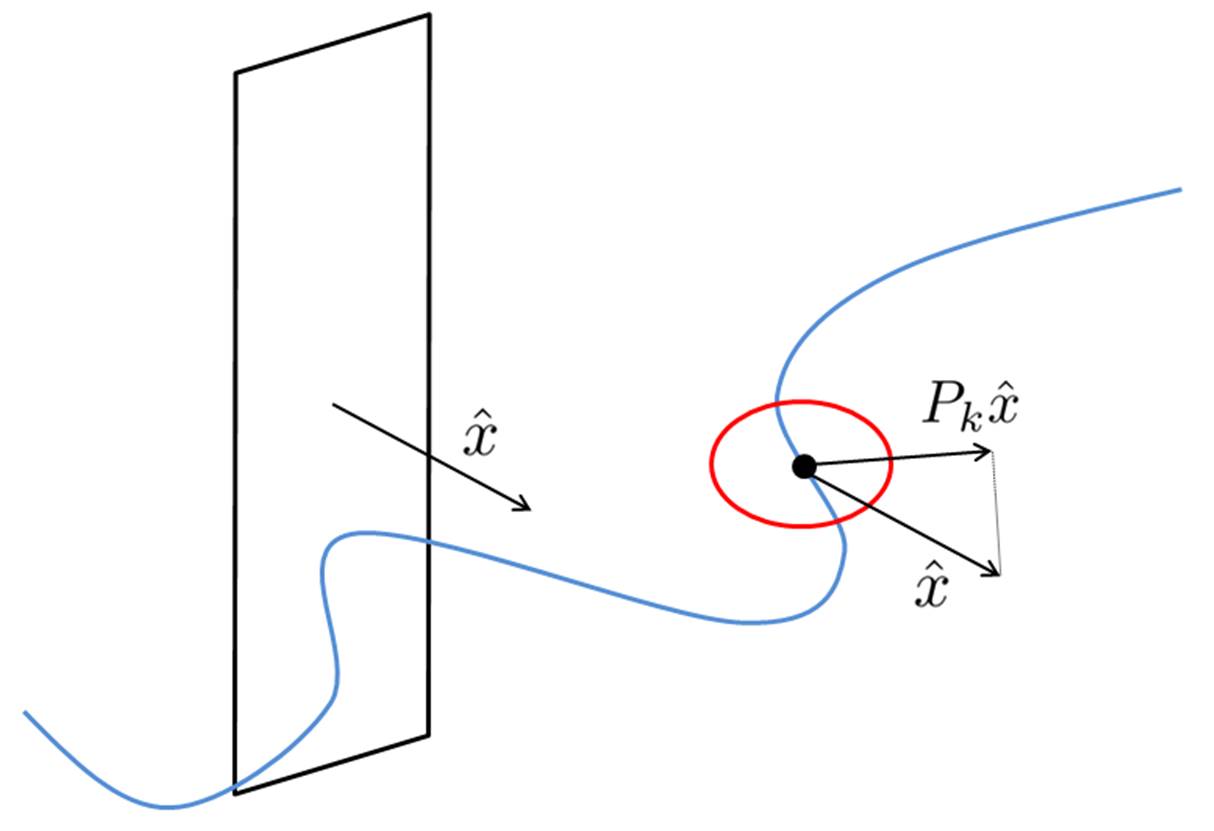}
 \caption{A visualization of the problem at hand.  The plane represents the partitioning surface, with normal $\hat{x}$.  In this case, we show a nodal line in three dimensions, which has $q=2$.  At each point on this line, we find the appropriate BW velocity using the projection $P_k\hat{x}$ of the unit vector $\hat{x}$ onto the normal plane of the line at that point.}
 \label{fig:codimension}
 \end{figure}

\subsection{General Codimension}

All of the above results have been derived for a standard codimension one Fermi surface, which is effectively a surface of quasi-one-dimensional modes, but the general idea can be carried over to the general codimension case with little trouble.  For a Fermi surface of codimension $q$, the effective theory will be described by a set of quasi-$q$-dimensional relativistic theories.  Suppose we have a specified codimension $q$ Fermi surface ($i.e.$ a surface of dimension $d-q$ embedded in $d$ dimensions).  The dispersion is generically linear around this surface.  In close analogy with the codimension one case, each patch on this surface has its own independent $q$-dimensional gapless field theory, which we take to be isotropic, so that each patch is characterized by a single Fermi velocity.\cite{foot4}  We do, however, allow the velocity to vary over the surface.  Unlike the codimension one case, we don't just have one normal direction to the Fermi surface, but rather a $q$-dimensional normal surface within which we have linear dispersion (see Figure \ref{fig:codimension}).  Within each patch of the Fermi surface, we can write an effective theory in terms of an independent Dirac field, yielding a Hamiltonian of the form:
\begin{equation}
H = \sum_{\textrm{patches}} \int d^dx\,v_F \overline{\Psi}_p (\gamma\cdot \partial_{x_\perp})\Psi_p
\end{equation}
for an appropriate set of $\gamma$ matrices.  The symbol $x_\perp$ represents the $q$ spatial coordinates corresponding to the normal directions at each patch.  Now choose a basis for $x_\perp$ which includes the direction of the projection of $\hat{x}_1$ into the normal surface (see Figure \ref{fig:codimension}).  It is now not hard to see that the appropriate BW velocity relating $x_0$ and $x_1$ will once again have a $\cos\theta$ factor, where now the angle is that between $\hat{x}_1$ and its projection onto the normal surface.  Since $\hat{x}_1$ is a unit vector, this factor can be written more succinctly as $|P_k\hat{x}_1|$, $i.e.$ the magnitude of the projection.  (Note that the projector is different for each patch of the Fermi surface, so it has $k$ dependence.)  The appropriate velocity to use in the Bisognano-Wichmann theorem will then be $v_F|P_k\hat{x}_1|$.  The appropriate temperature profile for each patch of the Fermi surface is then:
\begin{equation}
T_p(x) = \frac{v_F|P_k\hat{x}_1|}{2\pi x_1}
\end{equation}
The entanglement Hamiltonian is given by:
\begin{equation}
\tilde{H} = \sum_{\textrm{patches}}\int_{x_1>0} d^dx\frac{2\pi x_1}{v_F|P_k\hat{x}_1|} \overline{\Psi}_p(\gamma\cdot\partial_{x_\perp})\Psi_p
\end{equation}
To put these results to use, we need to know the entropy density contributed by each patch of the Fermi surface.  This is a fairly straightforward thermodynamics problem.  The thermal energy density per patch area at temperature $T$ is given by:
\begin{align}
\begin{split}
e(T) = \frac{1}{(2\pi)^d} \int d^qk \frac{v_F|k|}{e^{v_F|k|/T}+1} = \\
\frac{2\pi^{q/2}(2^q-1)}{(2\pi)^d(2v_F)^q} \frac{\Gamma(q+1)}{\Gamma(q/2)}\zeta(q+1)T^{q+1}
\end{split}
\end{align}
Making use of $de = Tds$, we will have $s(T) = \frac{q+1}{q}(e/T)$, or with some simplification:
\begin{align}
\begin{split}
s(T) = \frac{\pi^{q/2}(2^q-1)}{(2\pi)^d 2^q}& \frac{\Gamma(q+2)}{\Gamma(\frac{q}{2}+1)} \zeta(q+1) \bigg(\frac{T}{v_F}\bigg)^q \\
&\equiv A(q) \bigg(\frac{T}{v_F}\bigg)^q
\end{split}
\end{align}
where this equation defines the shorthand $A(q)$.  Following the same procedure that we did in the codimension one case, the entanglement entropy for a generic partition then becomes:
\begin{align}
\begin{split}
S = A(q)\int_{F.S.}\int_{P.S.}&\int_{x_1>0} dx_1 \bigg(\frac{|P_k \hat{n}_r|}{2\pi x_1}\bigg)^q =\\
\frac{A(q)}{(2\pi)^q(q-1)a^{q-1}}&\int_{F.S.}\int_{P.S.} |P_k \hat{n}_r|^q \equiv \\
\frac{N(q)}{a^{q-1}}  &\int_{F.S.}\int_{P.S.} |P_k \hat{n}_r|^q
\label{widomgen}
\end{split}
\end{align}
where the integrals are over the $(d-q)$-dimensional Fermi surface and the $(d-1)$-dimensional partitioning surface, and $a$ represents a short-distance cutoff.  The numerical coefficient (which will not be terribly important) is given by:
\begin{equation}
N(q) = \frac{\pi^{q/2}(2^q-1)}{(2\pi)^{d+q} 2^q (q-1)} \frac{\Gamma(q+2)}{\Gamma(\frac{q}{2}+1)} \zeta(q+1)
\end{equation}
Equation \ref{widomgen} is the generalization of the Widom formula to the general codimension case, $q>1$.  The factor of $|P_k\hat{n}_r|^q$ is the natural generalization of the ``flux factor" $|\hat{n}_r\cdot \hat{n}_k|$ found in the codimension one case.

There are a few notable features to this formula.  First of all, assuming $q>1$, there is no large distance divergence to the integrals, and the area law is strictly obeyed.\cite{foot5}  Also, the short-distance cutoff $a$ appears in the prefactor, making the overall magnitude non-universal.  Rather, it is only the relative magnitude of different choices of partitioning surfaces which is meaningful, as we will discuss below.  In the $q=1$ case, the dimensions of the Fermi surface, represented by $k_F$, provided all the necessary dimensionful factors, with the prefactor of the logarithm scaling as $(Lk_F)^{d-1}$.  For general $q$, the scaling goes as $L^{d-1}k_F^{d-q}/a^{q-1}$.  Another notable aspect is the generalized flux factor, $|P_k \hat{n}_r|^q$.  Since $|P_k \hat{n}_r| < 1$, this means that there is much less contribution to the entanglement entropy from patches of the Fermi surface that don't ``align" with the partitioning surface, in the sense of $\hat{n}_r$ lying in the normal surface.  And since $q$ appears as an exponent, we see that the higher codimension case is even more sensitive to these misalignments than the usual Widom formula.

\subsection{Nodal Line Entanglement Entropy}

We now specialize the results of the previous section to the case of Fermi nodal lines in three spatial dimensions, which have $q=2$.  Let the nodal line have a unit tangent vector $\hat{t}$.  The projection $P_k\hat{n}_r$ can then be written as $\hat{n}_r - (\hat{t}\cdot \hat{n}_r)\hat{n}_r$.  The entanglement entropy then takes the form:
\begin{equation}
\begin{split}
S = \frac{N}{a}&\int_{P.S.}\int_{N.L.} |\hat{n}_r - (\hat{t}\cdot \hat{n}_r)\hat{n}_r|^2\,\,\, = \\
\frac{N}{a}&\int_{P.S.}\int_{N.L.} (1 - (\hat{t}\cdot\hat{n}_r)^2)
\end{split}
\end{equation}
where the inner integral is over the nodal line.  This can equivalently be written in the form:
\begin{equation}
S = \frac{N}{a}\int_{P.S.}\int_{N.L.} |\hat{t}\times\hat{n}_r|^2
\end{equation}
Either of these equivalent forms can be regarded as the Widom formula for nodal lines.  Unlike the case of a codimension one Fermi surface, this result will be proportional to the size of the partitioning surface, without any logarithmic factors, so the area law for entanglement entropy will be obeyed.  Also, we note the explicit presence of the lattice scale $a$, making the coefficient non-universal.  But do not despair!  While the overall magnitude of the entanglement entropy for a given partition is not particularly meaningful, there is still important information contained in its shape dependence.  We can examine the \emph{relative} magnitude for different orientations and shapes of the partitioning surface.  The non-universal prefactors will cancel out of such ratios.  While the entanglement entropy does not tell us the overall size of the nodal line, it does allow us to map out the shape of the line.  For example, suppose we choose a planar partitioning surface, breaking the system up into two half-spaces.  Let the area of the plane (dictated by the system size) be $A$, and let it have fixed normal $\hat{n}$.  The resulting entanglement entropy is then:
\begin{equation}
S = \frac{NA}{a}\int_{N.L.}|\hat{t}\times\hat{n}|^2
\label{entplane}
\end{equation}
The integral is similar to, but not identical to, the length of the projection of the line onto the partitioning plane, which would be $\int |\hat{t}\times\hat{n}|$.  While the present integral is not directly equal to the projective size, like the codimension one case, it is a proxy for it.

\begin{figure}[t!]
 \centering
 \includegraphics[scale=0.4]{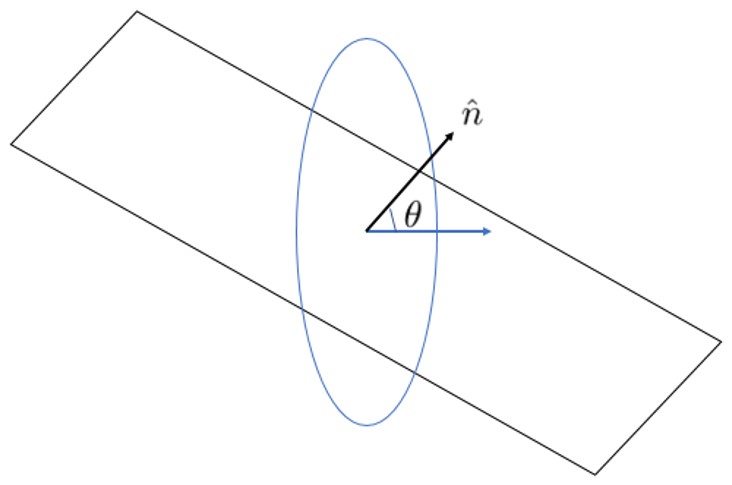}
 \caption{We consider a circular nodal line and a planar partitioning surface which is misaligned with it by angle $\theta$.  The entanglement entropy is greatest when the two normal vectors line up, $\theta = 0$.}
 \label{fig:circle}
 \end{figure}

As a concrete example, let us take our nodal line to be circular in shape, lying in a plane, as illustrated in Figure \ref{fig:circle}.  (This is not extraordinarily unrealistic.  Some form of symmetry protection may cause the nodal line to exist in a specific plane.  Also, the presence of interactions may smooth out the line under the renormalization group, resulting in a shape of high symmetry at low energies.)  Let the nodal circle lie in the $xy$ plane.  Without loss of generality, we then let $\hat{n}$ lie in the $xz$ plane, $\hat{n} = (\sin\theta,0,\cos\theta)$, with $\theta$ representing the angle between $\hat{n}$ and the normal to the circle.  Taking $\phi$ as the azimuthal angle in the plane, we have $\hat{t} = (-\sin\phi,\cos\phi,0)$.  The entanglement entropy is then:
\begin{align}
\begin{split}
S(\theta) = \frac{NA}{a}\int_0^{2\pi} &d\phi (\cos^2\theta + \sin^2\theta\cos^2\phi) = \\
&\frac{\pi NA}{a}(1+\cos^2\theta)
\end{split}
\end{align}
We therefore see that $S$ is maximized when $\theta = 0$, $i.e.$ when the nodal line lies in the same plane as the partitioning surface, and is minimized when it is perpendicular to the partitioning surface.  In particular, we can make the prediction that the entanglement entropy will be precisely half as large in the latter case, $S(\pi/2)/S(0) = 1/2$.  This is similar to the projective length $\ell(\theta) = \int |\hat{t}\times\hat{n}|$, which would obey $\ell(\pi/2)/\ell(0) = 1/\pi$.  We see that the planar entanglement entropy is highly anisotropic, as might be expected.  Moreover, a given shape will have a very specific form.  This provides a useful tool for mapping out the shape of a nodal line, purely based on ground state entanglement properties.  By examining the angular dependence of planar partitions, and comparing against the results for specific nodal line shapes, one can thereby get a general picture of the overall shape of the nodal line.

However, there are likely other systems which also have such anisotropic entanglement.  It would be nice to have a direct test which can verify that the anisotropy is indeed coming from a nodal line, as opposed to some other source.  Luckily, Equation \ref{entplane} for the planar entanglement entropy obeys a nontrivial relation which will not be obeyed by a generic anisotropic entropy source.  Let us look at three separate planar partitions, corresponding to three mutually orthogonal normal vectors $\hat{n}_1$, $\hat{n}_2$, and $\hat{n}_3$ (letting all planes have the same area $A$).  The sum of the entanglement entropies for these three partitions will obey:
\begin{equation}
\begin{split}
S_1+S_2+S_3 = &\frac{NA}{a}\int_{N.L.}\sum_{\hat{n}_1,\hat{n}_2,\hat{n}_3}\epsilon^{ijk}\hat{t}_j\hat{n}_k\epsilon^{i\ell m}\hat{t}_\ell\hat{n}_m = \\
&\frac{NA}{a}\int_{N.L.}2 = \frac{2NA}{a}\ell
\end{split}
\end{equation}
where $\ell$ is the length of the nodal line, and we have taken advantage of the completeness of the three vectors as a basis.  This relation says that, for an arbitrary orthogonal basis $\{\hat{n}_1,\hat{n}_2,\hat{n}_3\}$, the sum of the three entropies is always equal to the same constant, dictated by the length of the line.  This result is not satisfied by a generic anisotropic entropy source, and is therefore a unique distinguishing feature of nodal line anisotropy.  By verifying this relation, joined with an analysis of the precise form of the anisotropy, the entanglement entropy thereby provides us with a useful tool for determining the existence and shape of a Fermi nodal line.

\section{Conclusion}

In this work, we have generalized the Widom formula for Fermi surface entanglement entropy to the case of arbitrary codimension, by means of repurposing a trick for relativistic theories.  Along the way, we found an entanglement Hamiltonian for Fermi surface systems and a new derivation of the standard Widom formula.  Most importantly, we have found a Widom formula for nodal lines, which depends sensitively on the shape of the nodal line and its orientation relative to the chosen partitioning surface.  While the overall magnitude of the entanglement entropy is non-universal, the generalized Widom formula displays a characteristic angular dependence.  By comparing the entanglement entropy for different partitions, one can map out the shape of the nodal line, allowing for the detection of a nodal line purely based on ground state entanglement properties.  There is compelling evidence that nodal lines can exist in realistic systems, and entanglement entropy will serve as a useful tool for diagnosing these phases of matter.

\section*{Acknowledgments}

I would like to thank Ari Turner, Brian Swingle, Yahui Zhang, and Senthil Todadri for useful discussions.  In particular, I thank Ari Turner for sharing his ideas on Fermi surface entanglement Hamiltonians, Brian Swingle for encouragement during the early stages of this project, and Yahui Zhang and Senthil Todadri for sharing and discussing their work on nodal line spin liquids.  This work was supported by NSF DMR-1305741.

\section*{Appendix A: The Bisognano-Wichmann Theorem}

 \begin{figure}[t!]
 \includegraphics[scale=0.75]{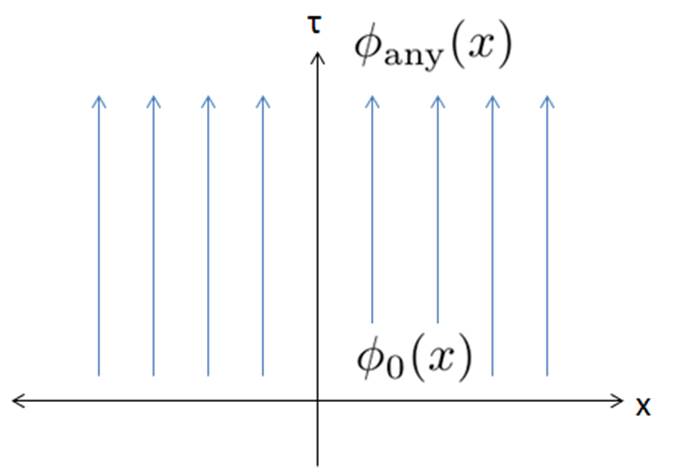}
 \caption{The wavefunction can be understood as a transition amplitude from $\phi_0$ to $\phi_\textrm{any}$ in imaginary time.}
 \label{fig:vertical}
 \end{figure}
 
 \begin{figure}[t!]
 \includegraphics[scale=0.5]{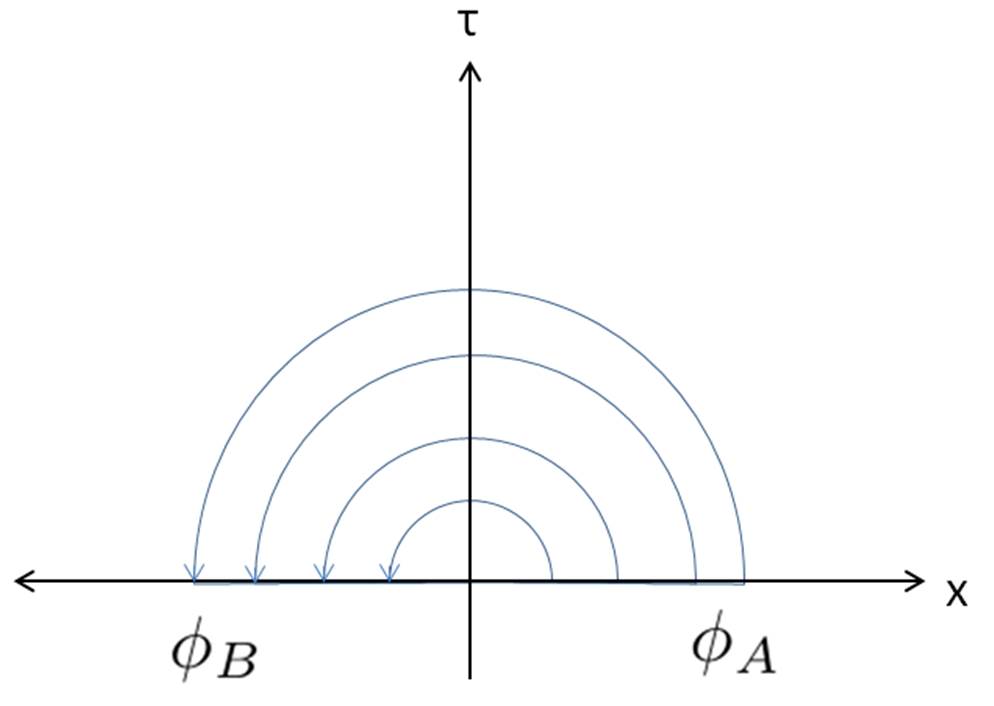}
 \caption{Equivalently, it can be understood as a transition amplitude from $\phi_A$ to $\phi_B$ in an angular time variable.}
 \label{fig:polar}
 \end{figure}

We review a quick argument for the Bisognano-Wichmann theorem for entanglement Hamiltonians in relativistic systems.  Let $\Psi_0[\phi] = \langle\phi |\Psi_0\rangle$ be the ground state of our system, as a function of $\phi$, schematically representing all fields in the theory.  We take advantage of the fact that $|\Psi_0\rangle$ can be obtained starting with a state corresponding to an arbitrary field configuration, $|\phi_{any}\rangle$, then evolving it in imaginary time, $e^{-H\tau}|\phi_{any}\rangle$.  As $\tau \rightarrow\infty$, the true ground state should be recovered (provided $|\phi_\textrm{any}\rangle$ has at least some nonzero overlap with the ground state, as a generic state does).  Thus, we can write the ground state as:
\begin{equation}
\langle\phi_0(x)|\Psi_0\rangle \propto \lim_{\tau\to\infty} \langle\phi_0(x)|e^{- H\tau}|\phi_{\textrm{any}}\rangle
\end{equation}
The above formula represents the amplitude for the state $|\phi_0(x)\rangle$ at imaginary time $\tau = 0$ to transition to the state $|\phi_{\textrm{any}}\rangle$ at imaginary time $\tau = \infty$.  In path integral language, this amplitude can be rewritten as:
\begin{equation}
\langle\phi_0(x)|\Psi_0\rangle \propto \int \mathcal{D}\phi' e^{\int_0^\infty d\tau \mathcal{L}_E[\phi']}
\end{equation}
where $\mathcal{L}_E$ is the Lagrangian expressed in imaginary time, and the path integral is over all configurations $\phi'(x,\tau)$ subject to the boundary conditions $\phi'(x,0) = \phi_0(x)$ and $\phi'(x,\infty) = \phi_{\textrm{any}}(x)$.  Then we go to polar coordinates in the $x_1\tau$ upper half-plane instead of Cartesian ones.  We can equally well regard this amplitude as that for the propagation from the configuration $\phi_A$ for $x_1>0$ to the configuration $\phi_B$ for $x_1<0$, with the angular coordinate taking the role of the time variable.  (See Figures \ref{fig:vertical} and \ref{fig:polar}).  Since the theory is relativistic, we can simply write this amplitude as:
\begin{equation}
\langle\phi_0(x)|\Psi_0\rangle \propto \langle \phi_A|e^{-\pi H_R}|\phi_B\rangle
\end{equation}
where $H_R = \frac{1}{c} \int_{x_1>0}d^dx\,x_1\mathcal{H}$ is the generator of boosts in the $x_1$ direction ($x_1\tau$ rotations).  (If the theory were not relativistic, then $H_R$ would not be a conserved quantity along the trajectory, necessitating a sort of ``time-ordering.")  The ground state density matrix for the system is then given by:
\begin{equation}
\rho_0(\phi,\phi') \propto \langle\phi_A|e^{-\pi H_R}|\phi_B\rangle\langle\phi_B'|e^{-\pi H_R}|\phi_A'\rangle
\end{equation}
and the reduced density matrix for region $A$ ($x_1>0$) is given by: 
\begin{equation}
\rho_A(\phi_A,\phi_A') \propto  \langle\phi_A|e^{-2\pi H_R}|\phi_A'\rangle
\end{equation}
Thus, up to an additive constant, we may write:
\begin{equation}
\tilde{H} = -\log\rho = \int_{x_1>0} d^dx\bigg(\frac{2\pi x_1}{c}\bigg)\mathcal{H}
\end{equation}
which is what we set out to show.

\section*{Appendix B: Connecting with Microscopic Variables}

As we have seen in the main text, the entanglement Hamiltonian takes a local form with respect to suitably chosen low-energy patch variables.  However, this does not imply that it is local with respect to the original microscopic fermions.  In this section, we will make connection with the microscopic fermions by employing an alternate technique for calculating the entanglement Hamiltonian.  We will find that indeed it is nonlocal with respect to microscopic variables, but in a fairly simple way.  The treatment here has been heavily influenced by a discussion with Ari Turner, who put forward the main logic of this method.

The important insight is to regard the higher-dimensional Fermi surface as a collection of one-dimensional Fermi surfaces.  To illustrate the principle, we will work through a two-dimensional model.  (Qualitatively similar results will hold in higher dimensions.) We take a Hamiltonian of the form:
\begin{equation}
H = \int dxdy \overline{\Psi}(-(\partial_x^2 + \partial_y^2) - k_F^2)\Psi
\end{equation}
modeling fermions at chemical potential $\mu = k_F^2$ just above the bottom of a band, where the dispersion is $\omega = k^2$ (setting the band mass to $1/2$ for simplicity).  Suppose $x$ is the normal direction to our partitioning cut.  We then Fourier transform in the transverse ($y$) direction:
\begin{equation}
H = \int \frac{dk_y}{2\pi} \int dx \Psi_{k_y}(-\partial_x^2 - (k_F^2 -k_y^2))\Psi_{k_y}
\label{1d}
\end{equation}
We have written $k_y$ as a subscript on the field $\Psi_{k_y}$ to emphasize that we can now treat it as a parameter.  For each fixed $k_y$, we have a theory of one-dimensional fermions with dispersion $\omega= k_x^2$ and chemical potential given by $\mu(k_y) = k_F^2-k_y^2$.  For $k_y > k_F$ ($i.e.$ for the region outside the original Fermi surface), the one-dimensional system is (trivially) gapped and will not contribute significantly to entanglement.  We can safely ignore this region without losing any interesting physics.  We therefore need to find the entanglement Hamiltonian for $k_y < k_F$, where we have a one-dimensional Fermi surface, consisting of two points.

We note the following important fact about these one-dimensional theories: while the full theory does not have any relativistic invariance, the low-energy description does, with velocity given by the effective Fermi velocity, $v_F(k_y) = 2\sqrt{k_F^2-k_y^2}$.  When a theory with Hamiltonian density $\mathcal{H}$ has relativistic invariance at speed $v$, we can write its entanglement Hamiltonian (for the half-space) as $\tilde{H} = \frac{2\pi}{v}\int_{x_1>0} d^dx\,x_1\mathcal{H}$.  Since our theory does not have full relativistic invariance, such a replacement will not give the exact entanglement Hamiltonian for the theory.  It will, however, give the correct contribution from low energy modes, which is the dominant contribution.  We can therefore make the rough approximation of using the BW result on each one-dimensional theory, resulting in the following entanglement Hamiltonian:
\begin{align}
\begin{split}
\tilde{H} = \int_{-k_F}^{k_F} \frac{dk_y}{2\pi} \frac{\pi}{\sqrt{k_F^2-k_y^2}}& \int dx\,x\bigg(\\
\Psi_{k_y}(-\partial_x^2 &- (k_F^2 -k_y^2))\Psi_{k_y}\bigg)
\end{split}
\end{align}
We can now explicitly perform the inverse Fourier transformation.  Consider the first term:
\begin{equation}
\begin{split}
\int &dx\,x \int dydy' \bigg(\\
\int_{-k_F}^{k_F}& \frac{dk_y}{2\pi} \frac{\pi}{\sqrt{k_F^2-k_y^2}} e^{ik_y(y'-y)}\bigg) \overline{\Psi}(x,y') (-\partial_x^2) \Psi(x,y) =
\label{fourier} \\
\int dx\,x &\int dydy' \bigg( \frac{\pi}{2} J_0(k_F(y'-y)) \bigg) \overline{\Psi}(x,y') (-\partial_x^2) \Psi(x,y)
\end{split}
\end{equation}
where $J_0$ is a Bessel function.  A similar analysis on the other term yields the full form of the entanglement Hamiltonian as:
\begin{align}
\begin{split}
\tilde{H} = \frac{\pi}{2}\int& dxdydy'\,x\bigg(\\
&J_0(k_F(y'-y)) \overline{\Psi}(x,y') (-\partial_x^2) \Psi(x,y) - \\
&\frac{k_F}{(y'-y)}J_1(k_F(y'-y)) \overline{\Psi}(x,y') \Psi(x,y)\bigg)
\label{fullent}
\end{split}
\end{align}
When the separation $y'-y$ is large, we can use asymptotic expressions for the Bessel functions.  In this limit, the quantity in parentheses above becomes:
\begin{align}
\begin{split}
\cos(k_F (y'-y))&\bigg(\\
&\sqrt{\frac{2}{\pi k_F (y'-y)}} \overline{\Psi}(x,y') (-\partial_x^2) \Psi(x,y) \\
&-  \frac{\sqrt{2k_F/\pi}}{(y'-y)^{3/2}} \overline{\Psi}(x,y') \Psi(x,y)\bigg)
\end{split}
\end{align}
We note the power law decay and also the oscillations at wavevector $k_F$, reminiscent of Friedel oscillations.  Just as in Friedel oscillations, the long-range physics here is due to the sharp change in occupation at the Fermi level.  We are essentially taking the Fourier transform of a function with a kink at $k_F$.  One can verify that the decay exponents $1/2$ and $3/2$ come directly from the behavior of the integral near $k_F$. The large separation behavior of the entanglement Hamiltonian, and thus its degree of nonlocality, is dictated entirely by physics near the Fermi surface.

The field $\Psi$ represents our microscopic fermionic variable, and the nonlocality of $\tilde{H}$ can be directly interpreted as the nonlocality of the hopping elements $h_{ij}$ of the quadratic form of the entanglement Hamiltonian.  As mentioned earlier, for any free fermion system on a lattice the entanglement Hamiltonian must take a quadratic form:
\begin{equation}
\rho \propto \exp(-\sum_{ij}h_{ij}c_i^\dagger c_j)
\end{equation}
The result obtained in Equation \ref{fullent} corresponds to the continuum limit of this expression, and we can directly see the nonlocality of the hopping elements.  The hopping element between sites will decay as $\cos(k_F(y'-y))/\sqrt{y'-y}$ at large separation in the $y$ direction.  The interaction in the $x$ direction is still local, so hopping should still be mostly nearest neighbor in that direction.


\begin{thebibliography}{9}

\bibitem{topo}  X.-G. Wen, \emph{Topological Order: From Long-Range Entangled Quantum Matter to an Unification of Light and Electrons}.  ISRN Condensed Matter Physics, Vol. 2013, 198710, arXiv:1210.1281v2 (2012)

\bibitem{area}  J. Eisert, M. Cramer, and M. B. Plenio, \emph{Area Laws for the Entanglement Entropy - A Review}.  Rev. Mod. Phys. 82 277, arXiv:0808.3773v4 (2008)

\bibitem{fermi1}  M. M. Wolf, \emph{Violation of the Entropic Area Law for Fermions}.  Phys. Rev. Lett.
96 010404, arXiv:quant-ph/0503219 (2005)

\bibitem{fermi2}  D. Gioev and I. Klich, \emph{Entanglement Entropy of Fermions in Any Dimension and the Widom Conjecture}.  Phys. Rev. Lett. 96, 100503, arXiv:quant-ph/0504151 (2005)

\bibitem{fermi3}  B. Swingle, \emph{Entanglement Entropy and the Fermi Surface}.  Phys. Rev. Lett. 105 050502, arXiv:0908.1724 (2009)

\bibitem{anushya}  A. Chandran, C. Laumann, and R. D. Sorkin, \emph{When is an Area Law not an Area Law?}.  Entropy 2016, 18(7), 240,  arXiv:1511.02996 (2015)

\bibitem{yang2}  H.-H. Lai, K. Yang, and N. E. Bonesteel, \emph{Violation of the Entanglement Area Law in Bosonic Systems with Bose Surfaces: Possible Application to Bose Metals}.  Phys. Rev. Lett. 111, 210402 (2013), arXiv:1306.2698v2

\bibitem{ramis}  R. Movassagh and P. Shor, \emph{Power law violation of the area law in quantum spin chains}.  Proceedings of the National Academy of Sciences (2016): 201605716, arXiv:1408.1657v3

\bibitem{yang1}  W. Ding, A. Seidel, and K. Yang, \emph{Entanglement Entropy of Fermi Liquids via Multidimensional Bosonization}.  Phys. Rev. X 2, 011012 (2012), arXiv:1110.3004v3

\bibitem{cfl}  R. V. Mishmash and O. I. Motrunich, \emph{Entanglement entropy of composite Fermi liquid states on the lattice: In support of the Widom formula}.  arXiv:1605.08787v2 (2016)

\bibitem{portal}  T. Grover, Y. Zhang, and A. Vishwanath, \emph{Entanglement Entropy as a Portal to the Physics of Quantum Spin Liquids}.   New J. Phys. 15 025002, arXiv:1302.0899 (2013)

\bibitem{foot1}
Actually, typically $k_F$ is of the same order as the inverse lattice spacing.  However, the point is that $k_F$ is part of the universal data characterizing a Fermi surface system, whereas the lattice spacing is not.

\bibitem{yang3}  H.-H. Lai and K. Yang, \emph{Probing critical surfaces in momentum space using real-space entanglement entropy: Bose versus Fermi}.  Phys. Rev. B 93, 121109(R) (2016) arXiv:1510.03428v2

\bibitem{balents}  A. A. Burkov, M. D. Hook, and L. Balents, \emph{Topological Nodal Semimetals}.  Phys. Rev. B 84, 235126, arXiv:1110.1089v2 (2011)

\bibitem{mullen}  K. Mullen, B. Uchoa, and D. T. Glatzhofer, \emph{Line of Dirac Nodes in Hyper-Honeycomb Lattices}.  Physical Review Letters 115 026403, arXiv:1408.5522v3 (2014)

\bibitem{weng}  H. Weng et al., \emph{Topological Node-Line Semimetal in Three Dimensional Graphene Networks}.  Phys. Rev. B 92 045108, arXiv:1411.2175v2 (2014)

\bibitem{liang}  C. Fang et al., \emph{Topological Nodal Line Semimetals with and without Spin-Orbital Coupling}.  Phys. Rev. B 92 081201, arXiv:1506.03449v3 (2015)

\bibitem{fang}  C. Fang et al., \emph{Topological Nodal Line Semimetals}.  Chinese Phys. B 25, 117106 (2016)  arXiv:1609.05414

\bibitem{fstop}  Y. X. Zhao and Z. D. Wang, \emph{Topological Classification and Stability of Fermi Surfaces}.  Phys. Rev. Lett. 110 240404, arXiv:1211.7241v3 (2013)

\bibitem{hasan}  G. Bian et al., \emph{Topological Nodal-Line Fermions in Spin-Orbit Metal $PbTaSe_2$}.  Nature Communications 7, 10556 (2016); G. Bian et al., \emph{Topological Nodal-Line Fermions in the Non-Centrosymmetric Superconductor Compound PbTaSe$_2$}.  arXiv:1505.03069 (2015)

\bibitem{hf1}  K. Izawa et al., \emph{Line Nodes in the Superconducting Gap Function of Noncentrosymmetric CePt$_3$Si}.  Phys. Rev. Lett. 94 197002, arXiv:cond-mat/0504350 (2005)

\bibitem{hf2}  I. Bonalde, W. Bramer-Escamilla, and E. Bauer, \emph{Evidence for Line Nodes in the Superconducting Energy Gap of Noncentrosymmetric CePt$_3$Si from Magnetic Penetration Depth Measurements}.  Phys. Rev. Lett. 94 207002 (2005)

\bibitem{yahui}  Y. Zhang and T. Senthil, \emph{$U(1)$ Nodal Line Spin Liquid}.  (Forthcoming)

\bibitem{peschel}
I. Peschel and V. Eisler, \emph{Reduced density matrices and entanglement entropy
in free lattice models}.  J. Phys. A: Math. Theor. 42 504003,  arXiv:0906.1663v3 (2009)

\bibitem{wichmann}
J. J. Bisognano and E. H. Wichmann, \emph{On The Duality
Condition For Quantum Fields}.  J. Math. Phys. 17, 303 (1976)

\bibitem{bianchi}
  E. Bianchi and R. C. Myers,
  \emph{On the Architecture of Spacetime Geometry}.  Class. Quantum Grav. 31 214002,  arXiv:1212.5183 (2012)
  
\bibitem{grover}  T. Grover, A. M. Turner, and A. Vishwanath, \emph{Entanglement Entropy of Gapped Phases and Topological Order in Three Dimensions}.  Phys. Rev. B 84 195120,  	arXiv:1108.4038 (2011)

\bibitem{temperature}
G. Wong, et al., \emph{Entanglement Temperature and Entanglement Entropy of Excited States}.  Journal of High Energy Physics 2013:20,  arXiv:1305.3291v2 (2013)

\bibitem{fradkin}
E. Fradkin, \emph{Field Theories of Condensed Matter Physics, Second Edition}.  Cambridge University Press, 2013.  Chapter 17.

\bibitem{cardy}  P. Calabrese and J. Cardy, \emph{Entanglement Entropy and Quantum Field Theory}.   	J. Stat. Mech. 0406:P06002,  arXiv:hep-th/0405152v3 (2004)

\bibitem{swingle}
B. Swingle and T. Senthil, \emph{Universal Crossovers Between Entanglement Entropy and Thermal Entropy}.  Phys. Rev. B. 87.045123,  arXiv:1112.1069 (2011)

\bibitem{mcgreevy}
B. Swingle and J. McGreevy, \emph{Area Law for Gapless States from Local Entanglement Thermodynamics}.  Phys. Rev. B 93, 205120 (2016) arXiv:1505.07106

\bibitem{topins}
L. Fidkowski, \emph{Entanglement Spectrum of Topological Insulators and Superconductors}.  Phys. Rev. Lett. 104.130502,  arXiv:0909.2654v4 (2009)

\bibitem{foot2}
We need to work with spin-$1/2$, as opposed to the simpler case of spinless fermions, in order to maintain the relativistic symmetry and use the BW theorem.  Since the fermions are free, we can find the result for spinless fermions by taking half of the obtained entanglement entropy for the spin-$1/2$ case.

\bibitem{foot3}
We relabel the momentum of one spin species $k_\perp\rightarrow -k_\perp$, in order to regard it as the time reverse of the other spin species.  This is simply done for convenience.

\bibitem{foot4}
Even if the theory started out anisotropic, interactions would likely drive the theory to an isotropic fixed point under the renormalization group, so this situation is actually fairly generic.  Also, as we shall see, the resulting entanglement entropy is purely geometric, and the details of the dispersion cancel out anyway.

\bibitem{foot5}
For $q>1$, the $x_1$ integral converges at large $x_1$.  Formally, if we had $q<1$, we would have a divergence needing to be cut off at $L$, leading to a power-law violation of the area law.  One might produce such a situation by looking at fractal Fermi surfaces\cite{fractal}, but the physical relevance of such models is unclear.

\bibitem{fractal}
L. Huijse and B. Swingle, \emph{Area law violations in a supersymmetric model}.  Phys. Rev. B 87, 035108,  arXiv:1202.2367 (2012)




\end{thebibliography}
\end{document}